\documentstyle[12pt, fleqn]{article}
\begin{document}
\renewcommand{\arraystretch}{2.0}
\baselineskip=0.6cm
\begin{titlepage}
\begin{center}
\large{\bf QCD Sum Rule Calculation for\\
   the Tensor Charge of the Nucleon } \\

\bigskip\bigskip
{\large Hanxin He$^{1,2,3}$ and Xiangdong Ji$^{1,2}$} \\
\bigskip
{\small\it $^1$ Center for Theoretical Physics, \\
Laboratory for Nuclear Science and Department of Physics}\\
{\small\it Massachusetts Institute of Technology, Cambridge, Mass 02139, USA}\\
\bigskip
{\small\it $^2$ Institute for Nuclear Theory}\\
{\small\it University of Washington}\\
{\small\it Seattle, Washington 98195, USA} \\
\bigskip
{\small\it $^3$ China Institute of Atomic Energy}\\
{\small\it P.O.Box 275(18), Beijing 102413, China }\\
{\small\it and Institute of Theoretical Physics, Academia Sinica, China}\\
\begin{center}
{\small (MIT-CTP-2551 ~~~ Hep-ph/9607XXX ~~~ July 1996)}
\end{center}

\bigskip
{\large\bf Abstract}
\end{center}
\par
The nucleon's tensor charges (isovector $g^v_T={\delta}u-{\delta}d$
and isoscalar $g^s_T={\delta}u+{\delta}d$) are calculated using the
QCD sum rules in the presence of an external tensor field. In addition to the
standard quark and gluon condensates, new condensates described by
vacuum susceptibilities are induced by the external field. The latter
contributions to $g^v_T$ and $g^s_T$ are estimated to be small. 
After deriving some simplifying formulas, 
a detailed sum rule analysis yields 
$g^v_T=1.29 \pm 0.51$ and $g^s_T = 1.37 \pm 0.55$,  
or ${\delta}u=1.33 \pm 0.53$ and ${\delta}d = 0.04 \pm 0.02$
at the scale of 1 GeV$^2$. 
\medskip\bigskip
\par
   PACS number(s):12.38Lg, 14.20Dh, 11.55Hx
\end{titlepage}

\begin{center}
{\bf I. INTRODUCTION }
\end{center}
\bigskip
\par   
   Studying nucleon charges such as the baryon charge,  the axial
charge, and the tensor charge from the underlying theory of the strong
interactions,  quantum chromodynamics (QCD),  is obviously important 
to understand the properties of the nucleon as well as
the QCD at low energies. On one hand,  these nucleon charges, defined
as matrix elements of bilinear fields
($\bar {\psi}{\gamma}^\mu{\psi}$ for baryon charge, 
$\bar{\psi}{\gamma}^\mu{\gamma}_5{\psi}$ for axial charge, 
and $\bar{\psi}{\sigma}^{\mu\nu}{\psi}$ for tensor charge), 
are fundamental
observables characterizing properties of the nucleon. On the other hand, 
these nucleon charges are connected through deep-inelastic 
sum rules to the leading-twist nucleon
structure functions----the spin averaged structure
function $f_1(x)$,  the quark helicity
distribution $g_1(x)$[1] and the transversity distribution $h_1(x)$ [2, 3].
For example,  the first moment of $g_1(x)$ is
connected to the axial charge $g_A$ by the Bjorken sum
rule [1]. Recently,  Jaffe and Ji [3] showed that the first moment of
$h_1(x)$ is related to the tensor charge ${\delta}{\psi}({\psi}=u, d...)$ of
the nucleon through
\begin{equation}
{\int}^1_{-1}h_1(x)dx={\int}^1_0(h_1(x)-\bar{h}_1(x))dx={\delta}{\psi} \ .
\end{equation}
Thus, the tensor charge,  like the axial charge,  is one of 
the important nucleon observables.

  So far, there is yet no experimental data on 
$h_1(x)$ and on the tensor
charge because $h_1(x)$ does not manifest itself in 
inclusive deep inelastic process
and there are no fundamental
probes which couple directly to the tensor current.
A first measurement of $h_1(x)$ may be performed at HERA 
with semi-inclusive deep-inelastic scattering and at 
RHIC with transversely polarized Drell-Yan process [4], 
from which one may extract the value of the tensor
charge. Therefore, theoretical study of the tensor charge
of the nucleon in QCD is very important. 

In Ref. [5],  we sought to understand the physical significance of the tensor
charge and made estimates in the MIT bag model and the QCD sum rule approach.
The QCD sum rule approach [6] has been used extensively in the past
as a useful tool to study hadron masses, coupling constants,
form factors, etc. (see [7] for a
review). To evaluate the tensor charge
and other charges, one makes an
extension of the standard QCD sum rules [6] 
to avoid infrared divergences arising from zero-momentum transfer
in bilinear quark operators. 
There are several equivalent formulations of the
extension: the two-point correlation function in an external
field, i.e., the external field QCD sum rule approach [8],  and the
three-point function approach [9]. In Ref.[5],  we used the three-point
function approach to calculate the tensor charge, including operators up to
dimension 6.

   In the present paper we refine and extend our work in Ref. [5],
in particular, we now include operators of dimension 8. 
To simply the calculation, we use instead 
the QCD sum rule approach in external field [8].
The basic idea is this:
Consider propagation of the nucleon current in the presence of an
external tensor field $Z_{\mu\nu}$. The two-point
correlation function of the nucleon current is
calculated up to terms linearly proportional to
$Z_{\mu\nu}$, and the coefficient of which is the linear response
of the correlation function to the external $Z_{\mu\nu}$ field.
The linear response is first calculated in terms of 
vacuum properties through an operator product 
expansion(OPE) [10]. In addition to the standard vacuum 
condensates[6], there are new condensates induced by the 
field $Z_{\mu\nu}$. The latter represents the response of the
QCD vacuum to the external field and can be described by
vacuum susceptibilities. The susceptibility terms 
are exactly equivalent to the bi-local
contributions in the three-point
correlation function approach [5]. On the other hand,
the linear response can be expressed in terms of properties of
physical intermediate hadron states, 
where the tensor charge enters. By matching the
two calculations in the certain momentum range, 
we extract the tensor charge.

We first look at the sum rules in which there 
are no contributions of vacuum susceptibilities. 
By eliminating the nucleon coupling constant 
through the sum rule for mass\cite{k}, we obtain 
the simplifying formulas of the tensor charges, 
\begin{eqnarray}\label{motion}
\delta u & = & -\frac{4(2\pi)^2\langle\bar{q}q\rangle}{m^3_N}
(1-\frac{9m_o^2}{16m_N^2}),  \nonumber \\
\delta d & = & \frac{\langle{g^2_cG^2}\rangle}{36m_N^4}
\end{eqnarray}
at the scale $\mu^2=M_N^2$. 
[The leading-log evolution of the tensor charge
is 
$$
g^{\alpha}_T(\mu^2)=[{\alpha}(\mu^2)/{\alpha}
(\mu_0^2)]^{\frac{4}{33-2n_f}}g^\alpha_T(\mu^2_0)
$$
where indices ${\alpha}=v, s, u, d$ specify $g^u_T={\delta}u$, 
$g^d_T={\delta}d$, $g^v_T=\delta u - {\delta}d$, and $g^s_T=\delta u 
+ {\delta}d$, and ${n_f}$ is the flavor number.] 
Substituting ${\langle{\bar{q}q}\rangle}=-(240 {\rm MeV})^3$, 
 $\langle{g_c^2G^2}\rangle=0.47 {\rm GeV}^4, 
 m_0^2=-\langle\bar{q}g_c{\sigma}\cdot 
 Gq\rangle/\langle{\bar{q}q}\rangle$
$\approx{0.8 {\rm GeV}^2}$,  and $m_N = 0.94 {\rm GeV}$, 
we get ${\delta}u=1.29$,  ${\delta}d\approx0.02$,  $g_T^v=1.27$ and
$g^s_T=1.31$. The uncertainty from the vacuum condensates
alone is at the level of 20\%. In order to extract these tensor charges
with a better accuracy,  we analyze the sum rules in a more 
careful way. The result is consistent with the simplifying 
analysis presented above.

The paper is organized as follows. In Sec. II we derive the nucleon sum
rules in the presence of an external tensor field. 
The sum rules are analysed in Sec. III, where we first 
derive the simplifying formulas for the tensor charges,
then estimate the vacuum susceptibilities,
and finally make a detailed sum rule analysis.
The summary and conclusions are given in the last section.

\vfill
\bigskip\bigskip
\begin{center}
{\bf II. QCD Sum Rules for the Nucleon in An External Tensor Field}
\end{center}

\medskip
To derive the QCD sum rules, 
we consider the two-point correlation function in the presence of 
an  external constant tensor field $Z_{\mu\nu}$, 
\begin{equation}
\Pi(Z_{\mu\nu}, p)=i\int d^4xe^{ipx}
\langle 0{\mid}T(\eta(x)\bar \eta(0)){\mid}0\rangle_Z , 
\end{equation}
where $\eta$ is the interpolating current 
for a proton [11], 
\begin{equation}
{\eta}(x)={\epsilon}_{abc}[u_a^T(x)C{\gamma}_{\mu}u_b(x)]{\gamma}_5
{\gamma}^{\mu}d_c(x)\ \ , 
\end{equation}
\begin{equation}
\langle0{\mid}{\eta}(0){\mid}N(p)\rangle={\lambda}_Nv_N(p) \ \ , 
\end{equation}
where $a$, $b$ and $c$ are color indices, superscript 
{\sl T} means transpose,  {\sl C}
is the usual
charge-conjugation matrix, and $v_N(p)$ the nucleon spinor normalized to
$\bar{v}v=2m_N$. To construct the sum rule for tensor charges, 
we expand the correlation function (3) to the first order in $Z_{\mu\nu}$, 
\begin{equation}
\Pi(Z_{\mu\nu}, p)={\Pi}_0(p)+Z_{\mu\nu}{\Pi}^{\mu\nu}(p) + ..., 
\end{equation}
where $\Pi_0$ is the correlation function which can be used to 
construct the nucleon-mass sum rules. In the following, we
focus on the linear response function $\Pi^{\mu\nu}(p)$.

The QCD sum rule is based on the principle of duality,  
which postulates the correspondence between a
description of correlation functions 
in terms of hadronic degrees of freedom (phenomenological side)
and that in quark and gluon degrees of freedom (QCD theory
side).
The phenomenological representation of the correlation function in an
external $Z_{\mu\nu}$ field can be expressed in terms of physical intermediate
states as
\begin{equation}
\Pi(Z_{\mu\nu}, p)={{{\lambda}^2_N}\over{{(p^2-m_N^2)}^2}}g_T^{\alpha}
Z_{\mu\nu}[\hat{p}{\sigma}^{\mu\nu}\hat{p}+m_N^2{\sigma}^{\mu\nu}
+m_N\{\hat{p}, {\sigma}^{\mu\nu}\}]+\cdots, 
\end{equation}
where the nucleon's tensor charge is defined by
\begin{equation}
\langle{N(p, s){\mid}J^{\alpha}_{\mu\nu}(0){\mid}N(p, s)}\rangle
=g_T^{\alpha} \bar{v}_N(p,s)\sigma_{\mu\nu}v_N(p,s) \ , 
\end{equation}
where $J^\alpha_{\mu\nu}=\bar{\psi}_\alpha\sigma_{\mu\nu}
\psi_\alpha$ with $\alpha = v, s, u,$  or $d$, 
specifying $J^{v(s)}_{\mu\nu}=\bar {u}\sigma_{\mu\nu}u\mp\bar{d}
\sigma_{\mu\nu}d$ and $J^{u(d)}_{\mu\nu}=\bar{u}{\sigma}_{\mu\nu}
u(\bar{d}{\sigma}_{\mu\nu}d)$. 

The theoretical side of the correlation function (3) can 
be computed in the deep Euclidean region using OPE. 
Matching the two results obtained above,  we 
can extract the tensor charges in terms of QCD  
parameters and vacuum condensates.

\vfill
\bigskip\bigskip
\noindent{\bf A. Operator Product Expansion}
\medskip

We now compute the OPE for the nucleon-current correlation function in an
external tensor field $Z_{\mu\nu}$. The coupling between
quarks and the external antisymmetric
tensor field $Z_{\mu\nu}$ is described by an additional term
\begin{equation}
\triangle{\cal L}=g_q{\bar q}{\sigma}^{\mu\nu}qZ_{\mu\nu}
\end{equation}
in QCD Lagrangian, where the coupling constant $g_q$
depends on the quark type as well as the field $Z_{\mu\nu}(g_q=g_u=-g_d$
for isovector type coupling,  $g_q=g_u=g_d$ for isoscalar type).
The equation of motion for quarks is now
\begin{equation}
(i\hat{D}+g_qZ_{\mu\nu}{\sigma}^{\mu\nu})q=0  \ \ , 
\end{equation}
where $D_{\mu}=\partial_\mu+ig_cA^a_\mu{{\lambda^a}\over2},  g_c$
is the QCD gauge coupling,  $\lambda^a$ are the Gell-Mann
matrices,  and current masses of up and down quarks are neglected.

To calculate the Wilson coefficients in OPE [10], we
construct the coordinate-space quark
propagator in the presence of the external
field $Z_{\mu\nu}$. Following the method of Refs.[8] and [12],  we find
\begin{eqnarray}
S^{ab}_{ij} & = & \langle0{\mid}Tq^a_i(x)\bar{q}^b_j(0){\mid}0\rangle \nonumber \\
   & = & \frac{i\delta^{ab}(\hat{x})_{ij}}{2\pi^2x^4}-
\frac{\delta^{ab}}{4\pi^2x^4}g_q[x^2Z_{\mu\nu}(\sigma^{\mu\nu})_{ij}-4x^\nu
x_\rho Z_{\mu\nu}(\sigma^{\mu\rho})_{ij}] \nonumber \\
 & & +{\frac{i}{32\pi^2x^2}}g_c{\frac{\lambda^n_{ab}}{2}}G^n_{\mu\nu}(\hat{x}
\sigma^{\mu\nu}+\sigma^{\mu\nu}\hat{x})_{ij} \nonumber \\
 & & -{1\over{12}}\delta^{ab}\delta_{ij}{\langle}{0}{\mid}\bar{q}q{\mid}0\rangle
-{\frac{\delta^{ab}}{24}}Z_{\mu\nu}g_q\chi\langle\bar{q}q\rangle
(\sigma^{\mu\nu})_{ij} \nonumber \\
 & & -i{\frac{\delta^{ab}}{48}}g_q\langle\bar{q}q
{\rangle}Z_{\mu\nu}(\hat{x}\sigma^{\mu\nu}
+\sigma^{\mu\nu}\hat{x})_{ij}+{\frac{\delta_{ab}}{192}}\delta_{ij}x^2\langle{0}
{\mid}\bar{q}g_c{\sigma}\cdot Gq{\mid}0\rangle \nonumber \\
 & & +{\frac{\delta^{ab}}{288}}g_q\langle\bar{q}q{\rangle}Z_{\mu\nu}[x^2(\kappa-
2\zeta)(\sigma^{\mu\nu})_{ij}+2x^{\nu}x_{\rho}(\kappa+\zeta)(\sigma^{\mu\rho})
_{ij}]+\cdots   .
\end{eqnarray}
The first three terms in Eq.(11) are the perturbative propagator
for a massless free quark and its interactions with the external 
$Z_{\mu\nu}$ field and the vacuum gluon field $G^n_{\mu\nu}$. 
The next five terms are non-perturbative, 
arising from the quark and the mixed quark-gluon
condensates, and the condensates induced by the 
external field $Z_{\mu\nu}$. In fact, due to 
breakdown of Lorentz invariance, the 
vacuum expectation values $\langle\bar{q}\sigma_{\mu\nu}q\rangle, 
 \langle\bar{q}g_c{\frac{\lambda^n}{2}}G^n_{\mu\nu}q\rangle, 
\langle\bar{q}g_c\gamma_5\widetilde{G}_{\mu\nu}q\rangle$, where
$\widetilde{G}_{\mu\nu}={\frac{1}{2}}\epsilon_{\mu\nu rs}
G^{rsn}{{\lambda^n}\over2}$, become non-zero. To characterize
these vacuum response, we define the 
induced susceptibilities $\chi, \kappa$,  and
$\zeta$ through
\begin{eqnarray}
\lefteqn{\langle{0}{\mid}\bar{q}\sigma_{\mu\nu}q{\mid}0\rangle_Z
=g_q\chi Z_{\mu\nu}\langle\bar{q}q\rangle \ ,}  \nonumber \\
\lefteqn{\langle 0{\mid}\bar{q}g_c{{\lambda^n}\over2}G^n_{\mu\nu}q{\mid}0\rangle_Z
=g_q\kappa Z_{\mu\nu}\langle\bar{q}q\rangle \ ,}  \nonumber \\
\lefteqn{\langle{0}{\mid}\bar{q}g_c\gamma_5\widetilde{G}_{\mu\nu}q{\mid}0\rangle_Z
=-ig_q\zeta Z_{\mu\nu}\langle\bar{q}q\rangle \ .} 
\end{eqnarray}

Using 
\begin{equation}
\langle{0}{\mid}T(\eta(x)\bar{\eta}(0)){\mid}0\rangle
=-2\epsilon^{abc}\epsilon^{a'b'c'}Tr\{S(x)^{bb'}_u\gamma_{\nu}C
S(x)^{aa'}_u C\gamma_{\mu}\}\gamma_5\gamma^{\mu}S(x)^{cc'}_d\gamma^\nu
\gamma_5 , 
\end{equation}
and Eq. (11), it is straightforward to carry out the OPE. 
The corresponding Feynman diagrams are shown in Figs. 1 and 2. The
results for the correlation function (3), including
the contributions up to dimension 8, is
\begin{equation}
\Pi(Z_{\mu\nu}, p)= Z_{\mu\nu}\Big[W_1\hat{p}\sigma^{\mu\nu}\hat{p}
+W_2\sigma^{\mu\nu}+W_3\{\hat{p}, \sigma^{\mu\nu}\}+\cdots \Big]
\end{equation}
where
\begin{eqnarray}
W_1 & = &-{\frac{g_d}{24\pi^2}}\chi\langle\bar{q}q\rangle{ln(-p^2)}
+{\frac{g_d}{16\pi^2p^2}}\zeta\langle\bar{q}q\rangle+{\frac{g_d}{576\pi^2
p^4}}\chi\langle\bar{q}q\rangle\langle{g^2_cG^2}\rangle \nonumber \\
& & +{\frac{2g_u}{p^4}}\langle\bar{q}q\rangle^2
-{\frac{3g_u}{2p^6}}\langle\bar{q}q\rangle\langle
\bar{q}g_c{\sigma}{\cdot}Gq\rangle \ \ ,  
\end{eqnarray}
\begin{eqnarray}
W_2 & =& {\frac{g_d}{32\pi^4}}p^4ln(-p^2)+{\frac{g_d}{24\pi^2}}(4\kappa+\zeta)
\langle\bar{q}q{\rangle}ln(-p^2) \nonumber \\
& & +{\frac{g_d}{384\pi^4}}\langle{g^2_c}G^2\rangle{ln(-p^2)}
-{\frac{2g_u}{3p^2}}\langle\bar{q}q\rangle^2    \ \ , 
\end{eqnarray}
\begin{eqnarray}
W_3 & = &{\frac{g_u}{2\pi^2}}\langle\bar{q}q\rangle{ln(-p^2)}
+{\frac{g_u}{24\pi^2p^2}}ln(-p^2/{\Lambda^2})\langle\bar{q}g_c{\sigma}
{\cdot}Gq\rangle \nonumber \\
& & -\frac{g_u}{3p^2}\chi\langle\bar{q}q\rangle^2+\frac{g_u}{24p^4}\chi
\langle\bar{q}q\rangle\langle\bar{q}g_c{\sigma}{\cdot}Gq\rangle \nonumber \\
& & -\frac{g_u}{18p^4}(\kappa+4\zeta)\langle\bar{q}q\rangle^2
-\frac{7g_u+g_d}{288\pi^2p^4}\langle\bar{q}q\rangle
\langle{g^2_cG^2}\rangle  \ \ .
\end{eqnarray}

\bigskip\bigskip
\noindent{\bf B. Sum Rules}
\medskip

The QCD sum rules are obtained by equating the OPE results in 
Eqs.(14)--(17) and the phenomenological description in Eq.(7), 
and performing Borel transformation. The renormalization scale 
dependence is also included properly. There are three different
Dirac structures: chiral-odd $Z_{\mu\nu}\hat{p}\sigma^{\mu\nu}\hat{p}$
and $Z_{\mu\nu}\sigma^{\mu\nu}$,  and chiral-even  $Z_{\mu\nu}\{\hat{p} , 
\sigma^{\mu\nu}\}$,  each of which can be used to construct a sum rule and
extract the tensor charges. Note that the phenomenological representation of
$\Pi(Z_{\mu\nu}, p)$ in Eq.(7) contains the double-pole 
as well as various single-pole contributions. The latter terms 
can be treated in several ways. Here we eliminate them 
by multiplying both
sides of the sum rules by $m^2_N-p^2$ before Borel transformation. 
We find the sum rule for the structure
$Z_{\mu\nu} \hat{p}\sigma^{\mu\nu}\hat{p}$, 
\begin{eqnarray}
\lefteqn{-\frac{g_d\chi{a}}{24L^{4/9}}M^2(m^2_NE_0-M^2E_1)
+\frac{g_d\zeta{a}}{16}m_N^2L^{4/27}-\frac{g_d\chi{ab}}{576L^{4/9}}
(1+\frac{m^2_N}{M^2})} \nonumber \\
\lefteqn{+\frac{g_u a^2}{2}(1+\frac{m^2_N}{M^2})L^{16/27}-\frac{3g_um_0^2a^2}
{8M^2}(1+\frac{m^2_N}{2M^2})L^{4/27}} \nonumber \\
\lefteqn{=g^{\alpha}_T(\mu_0){\beta}^2_Ne^{-m_N^2/M^2},}
\end{eqnarray}
for $Z_{\mu\nu}\sigma^{\mu\nu}$, 
\begin{eqnarray}
\lefteqn{-\frac{g_d}{4}M^6(m^2_NE_2-3M^2E_3)+{g_d\over{24}}(4\kappa+\zeta)aM^4
(m^2_NE_1-2M^2E_2)} \nonumber \\
\lefteqn{-\frac{g_d}{96}bM^2(m_N^2E_0-M^2E_1)+{\frac{g_u}{6}}a^2m_N^2} \nonumber \\
\lefteqn{=g^{\alpha}_T(\mu_0)m^2_N{\beta}^2_Ne^{-m^2_N/M^2},}
\end{eqnarray}
and for $Z_{\mu\nu}\{\hat{p}, {\sigma}^{\mu\nu}\}$, 
\begin{eqnarray}
\lefteqn{\frac{g_u a}{2}M^2(m_N^2E_0-M^2E_1)L^{4/27}+\frac{g_u m_N^2
m_0^2a}{24L^{8/27}}(ln(m_N^2/\Lambda^2)-1)} \nonumber \\
\lefteqn{-\frac{g_u}{96L^{4/9}}\chi m_0^2 a^2(1+\frac{m_N^2}{M^2})
-\frac{g_u}{72}(\kappa+4\zeta)a^2(1+\frac{m_N^2}{M^2})L^{16/27}} \nonumber \\
\lefteqn{+\frac{g_u}{12}\chi m_N^2 a^2
+\frac{1}{288}(7g_u +g_d)ab(1+\frac{m_N^2}{M^2})L^{4/27}} \nonumber \\
\lefteqn{=g_T^{\alpha}(\mu_0)m_N{\beta}^2_Ne^{-m^2_N/M^2}.}
\end{eqnarray}
Here $a=-(2\pi)^2\langle\bar{q}q\rangle, \ b=\langle g_c^2G^2\rangle, \
m_0^2a=(2\pi)^2\langle\bar{q}g_c{\sigma}\cdot Gq\rangle, \ {\beta}^2_N=(2\pi)^4
{\lambda}_N^2/4$, and $M$ is the Borel mass.
The factors $E_n=1-{\sum}{\frac
{x^n}{n!}}e^{-x}$ with $x=s_0/M^2$ account for the sum of the
contributions from excited states,  where $s_0$ is an effective continuum
 threshold. The anomalous dimension of
various operators, including that of the tensor current, is taken into accout
through the factor $L=ln(M^2/{\Lambda^2})/ln({\mu}^2_0/{\Lambda^2})$ \ 
[6, 11],  where $\mu_0$ is the inital renormalization 
scale (500 MeV) and $\Lambda=\Lambda_{QCD}$
 is the QCD scale parameter (100 MeV). The tensor
charges $g_T^{\alpha}$ in Eqs.(18)-(20) are renormalized at the 
scale $\mu_0$.

\bigskip\bigskip
\begin{center}
{\bf III. Sum Rule Analysis}
\end{center}
\medskip\medskip

Let us now analyse the sum rules derived in the previous sections and
extract the tensor charges $g^v_T,  g^s_T$ and $\delta u,  \delta d$.
In principle, one expects to obtain the same result 
from each of the sum rules in Eqs.(18)-(20). In practice,  however, 
some sum rules work better than the others, as was evident in
similar studies [8, 14, 15]. In our case, the sum rule in Eq. (18) 
from the structure $\hat{p}\sigma^{\mu\nu}\hat{p}$ is preferred 
for the following reasons. Firstly,  
from Eqs.(14)-(17) the structure $\hat{p}\sigma^{\mu\nu}\hat{p}$ contains
extra powers of momentum in the numerator compared with structures $\sigma^
{\mu\nu}$ and $\{\hat{p},  \sigma^{\mu\nu}\}$. Hence, the 
resulting sum rule converges better after Borel transformation. 
Secondly, the sum rule in Eq. (18) is known better 
phenomenologically, because the terms involving the poorly-known
susceptibilities are numerically small compared with
the dominant dimension-6 and 8 terms. Hereafter, we 
discard the sum rule in Eq. (19). 

\vfill
\bigskip
{\bf A. Simple Formulas for the Tensor Charges}
\bigskip

Consider the sum rules (18) for the structure 
$\hat{p}\sigma^{\mu\nu}\hat{p}$. As a first 
approximation, we disregard the anomalous dimensions,  
let $M=m_N$, and use $\beta^2_Ne^{-m_N^2/M^2}{\mid}_{M^2=m^2_N}=-\pi^2
m_N^3\langle\bar{u}u\rangle$ from the simplified 
mass sum rules [11]. In the case of $g_u=1, g_d=0$, 
we find a simple formula for the tensor charge $\delta u$, 
\begin{equation}
\delta u =  -\frac{4(2\pi)^2\langle\bar{q}q\rangle}{m^3_N}
(1-\frac{9m_o^2}{16m_N^2}). 
\end{equation}
Similarly, from the sum rule (20) with the structure $\{\hat{p}, 
\sigma^{\mu\nu}\}$, we get 
\begin{equation}
\delta d  =  \frac{\langle{g^2_cG^2}\rangle}{36m_N^4} \ . 
\end{equation}

The result obtained above is independent of the
susceptibilities $\chi, \kappa$ and $\zeta$. Thus, one can 
get a simple estimate of the tensor charges without knowing
them. Substituting the vacuum condensates
into the above equations,  we find $\delta u\approx 1.29, 
\ \delta d\approx 0.02$, or $g_T^v\approx1.27$ and 
$g_T^s\approx1.31$ (at $\mu^2=m_N^2$). 
Interestingly, $\delta d $ is much smaller than $\delta u$, or 
isovector $g_T^v$ and isoscalar $g_T^s$ tensor charges
have similar sizes. 

\bigskip
{\bf B. Estimates of Susceptibilities}
\bigskip

To make a more detailed analysis, we need
the values of the susceptibilities $\chi, \kappa$,  and $\zeta$
defined in Eq.(12). These susceptibilities 
correspond to the bi-local contributions in the three-point correlation
function approach [5, 9]. In fact, using Eq. (9), it is easy to show
\begin{equation}
\langle{0}{\mid}\bar{q}{\sigma}^{\mu\nu}q{\mid}0\rangle_Z=g_qZ_{\mu\nu}
\lim_{p_\lambda\to 0}{i\over 6}{\int}d^4x e^{ipx}\langle 0{\mid}T(\bar{q}
{\sigma}_{\alpha\beta}q(x),  \bar{q}\sigma^{\alpha\beta}q){\mid}0\rangle\ \ .
\end{equation}
Combining with Eq.(12),  we obtain
\begin{equation}
\chi\langle\bar{q}q\rangle=\lim_{p_\lambda\to 0}{i\over 6}{\int}d^4x
e^{ipx}\langle 0{\mid}T(\bar{q}{\sigma}_{\alpha\beta}q(x),  \bar{q}
\sigma^{\alpha\beta}q){\mid}0\rangle
{\equiv}{1\over 6}\Pi_\chi(0)    \ \ , 
\end{equation}
where $\Pi_\chi(0)$ is a bi-local correlator defined in the three-point
correlation function approach (see Eq.(23) of Ref.[5]). Similarly, we have
\begin{equation}
\kappa\langle\bar{q}q\rangle={i\over 6}{\int}d^4x\langle 0{\mid}T(\bar{q}
g_c{\frac{\lambda^n}{2}}G^n_{\alpha\beta}q, \bar{q}\sigma^{\alpha\beta}q)
{\mid}0\rangle
{\equiv}{1\over 6}\Pi_{\kappa}(0)     \ \ ,  
\end{equation}
\begin{equation}
\zeta\langle\bar{q}q\rangle=-{1\over 6}{\int}d^4x\langle{0}{\mid}T(\bar{q}
g_c\gamma_5\widetilde{G}_{\alpha\beta}q, \bar{q}\sigma^{\alpha\beta}q){\mid}
0\rangle
\equiv-{1\over 6}\Pi_\zeta(0)              \ \ . 
\end{equation}
Substituting the above expressions for $\chi , \kappa$ and
$\zeta$ into Eqs.(15)-(17) and (18)-(20),  we 
obtain the same result obtained in the three-point correlation function
approach [5].

Eq.(24)-(26) provide 
useful formulas to determine the susceptibilities $\chi,  \kappa $ and $\zeta$.
To find $\chi$, we assume $\rho(1^{--})$ and $B(1^{+-})$ 
meson dominance in Eq. (23). Estimating the 
coupling constants of relevant currents with 
the meson states again using the QCD sum rules, we get [5]
$\Pi_\chi (0)\approx(0.15 {\rm GeV})^2$,  or
\begin{equation}
\chi {a}=-{\frac{(2\pi)^2}{6}}\Pi_\chi (0)\approx -0.15 {\rm GeV}^2.
\end{equation}
As pointed out in Ref.[5],  the small $\Pi_\chi(0)$ comes from the
cancellation of the two resonances. However, even if $\rho$
meson does not couple to the current $\bar{q}\sigma^{\mu\nu}q$ [13], 
we have $\Pi_\chi (0)\approx(0.21 {\rm GeV})^2$ and 
$\chi a\approx{-0.29 {\rm GeV}^2}$,  which are still small.

The susceptibilities $\kappa$ and $\zeta$ can be estimated 
in a similar way. To get just an order of magnitude,
we assume only B-meson couple to the current $\bar{q}
\sigma_{\mu\nu}q$. We find 
\begin{equation}
\Pi_\zeta(0)=-\Pi_{\kappa}(0)\approx e^{m_v^2/M^2}({\frac{2m_0^2}{m_v^2}}
\langle m\bar{q}q\rangle+{\frac{M^2}{8\pi^2m_v^2}}\langle g_c^2G^2\rangle)
\ \ , 
\end{equation}
where M and $m_v$ are the Borel mass and the vector meson mass, 
respectively. Substituting $\langle m\bar{q}q\rangle\approx-(100 {\rm MeV})^4, 
 \langle g_c^2G^2\rangle\approx 0.47 {\rm GeV}^4, 
m_v\approx 1.2 {\rm GeV},  M\approx 1 {\rm GeV}$, 
we get $\Pi_\zeta(0)=-\Pi_\kappa(0)\approx 0.017{\rm GeV}^4$, or
\begin{equation}
\zeta a=\kappa a\approx 0.10 {\rm GeV}^4   \ \ .
\end{equation}
The above estimates of $\chi,  \kappa$ and $\zeta$ are admittedly
crude. Fortunately, 
as we shall see from the anabelow,  the effects of susceptibilities
on the tensor charges $g^v_T$ and $g^s_T$ are small.

\bigskip\bigskip
{\bf C. Sum Rule Analysis}
\medskip\medskip

Now we perform a standard sum rule analysis for Eqs. (18) and (20).
Multiplying both sides of equations by $L^{-4/27}$,  
we study $g^\alpha_T(\mu)$ as function of Borel mass $M^2$.
In our calculation, $s_0$ is taken to be $2.3 {\rm GeV}^2$, 
$\beta_N ^2 = 0.26 {\rm GeV}^6$, and other parameters are the 
same as those following Eq. (2). The solutions for $\delta u(\mu^2)$
and $\delta d(\mu^2)$ are plotted as function of $M^2$ in Fig.3. 
$\delta u$ from the two sum rules are shown in solid and 
dashed lines, $\delta d$ are shown in dot-dashed and short-dashed 
lines. We take the spread of the two sum rule predictions 
as our theoretical uncertainty. Averaging the two results, we have 
\begin{equation}
 \delta u=1.33\pm 0.53 \ ,  \ \ \delta d=0.04 \pm 0.02
\end{equation}
at the scale $\mu^2=M^2=(1 {\rm GeV})^2$. 
Correspondingly, 
\begin{equation}
 g^v_T=1.29\pm 0.51 \ , \ \ g^s_T=1.37\pm 0.55 ,
\end{equation}
which are consistent with the simplified results early. 

\bigskip\bigskip
\begin{center}
{\bf IV. Summary and Conclusions}
\end{center}
\medskip\medskip

In this paper we studied the nucleon's tensor
charges (isovector one $g^v_T$,  isoscalar one $g^s_T$,  and
$\delta u$ and $\delta d$) by means of QCD sum rule
approach. The calculations include terms up to
dimension 8, which presumably gives more reliable 
predictions than our previous result. Our final numbers 
for the proton's tensor charges are
$g^v_T=1.29\pm 0.51, \ \ g^s_T=1.37\pm 0.51, \ \ \delta u=1.33\pm 0.53, 
\ \ \delta d=0.04\pm 0.02 $ at the scale of 1 GeV$^2$.

In obtaining the above results, we 
used different methods to analyze the sum rules
for $g_T^\alpha$. The first one is to derive the simplifying
formulas for the tensor charges at $\mu^2=m_N^2$. 
The second method
is to analyze the sum rules with different chiral structure,  from which we 
obtained the tensor charge at the scale of 1 GeV$^2$. 
Both analyses seem to be consistent. 

Our QCD sum rule calculation shows that 
$\delta u\gg \delta d$, which means that the up quark
dominates the contribution in a transversely-polarized proton. 
Furthemore, the isovector
tensor charge $g_T^v$ of the proton is comparable in magnitude 
with the isovector axial charge $g_A$. To the contrary, 
the isoscalar tensor charge is markedly different from the
isoscalar axial charge. We hope that
the future experimental 
measurement will test the present prediction. 

At last, we wish to point out that since $\delta d$ is small, 
the precise value of which, including its sign, is difficult
to calculate. Futhermore, the sum rules presented here 
have no stable solutions, which may reflect that some relevant physics
is missing in the OPE. For instance, instanton physics, 
important in the case of the nucleon mass
\cite{q}, might stablize these sum rules and improve 
the present prediction. 

\medskip
\par 
We thank Xuemin Jin for useful discussions. 
H.X. He was supported in part by the National Natural Science Foundation
 of China and the Nuclear Science Foundation of China. X. Ji was supported in
part by funds provided by the U.S. Department of Energy (DOE)
under cooperative agreement No.DF-FC02-94ER40818.

\vspace{1.5cm}
{\large Figure Captions}
\vspace{0.5cm}
\begin{description}
\item [Fig.1.] Diagrams for the calculation of the Wilson coefficients
of the correlation function correspnding to the structures
$\hat{p} \sigma^{\mu\nu}\hat{p}$ and $\sigma^{\mu\nu}$.
\item [Fig.2.] Diagrams for the calculation of the Wilson coefficients
 of the correlation function corresponding to the structure $\{{\hat{p}},
\sigma^{\mu\nu}\}$.
\item [Fig.3.] The solutions of tensor charges from 
Eqs. (18) and (20). The solid and dashed lines represent 
$\delta u$, and the dot-dashed and short-dashed lines 
represent $\delta d$, with upper curves from Eq. (18). 
\end{description}


\medskip
\begin{thebibliography}{efg}
\bibitem{a} J. D. Bjorken, \ \ {\em Phys. Rev.} {\bf 148}, 1467(1966);
 {\em Phys. Rev.} {\bf D1}, 1376(1971). 
\bibitem{b} J. Ralston and D. E. Soper {\em Nucl. Phys.} {\bf B152}, 109(1979).
\bibitem{c} R. L. Jaffe and X. Ji,  {\em Phy. Rev. Lett} {\bf 67}, 552(1991);
{\em Nucl. Phys.} {\bf B375},  527(1992)
\bibitem{d} The RSC proposal to RHIC,  1993;  and
the HERMES proposal to HERA,  1993.
\bibitem{e} H. X. He and X. Ji,  {\em Phys. Rev.} {\bf D52},  2960(1995).
\bibitem{f} M. A. Shifman,  A. J. Vainshtein,  and V. I. Zakharov, 
{\em Nucl. Phys.} {\bf B147},  385(1979); {\bf B147}, 448(1979).
\bibitem{g} L. J. Reinders,  H. R. Rubinstein,  and S. Yazaki,  
{\em Phys. Rep.} {\bf 127},  1(1985)
\bibitem{h} B. L. Ioffe and A. V. Smilga,  {\em Nucl. Phys.} {\bf B232}, 
109(1984).
\bibitem{i} V. M. Balisky and A. V. Yung,  {\em Phys. Lett.} {\bf B129}, 328(1983).
\bibitem{j} K. G. Wilson,  {\em Phys. Rev.} {\bf 179}, 1499(1969).
\bibitem{k} B. L. Ioffe,   {\em Nucl. Phys.} {\bf B188}, 317(1981);
V. M. Belyaev and B. L. Ioffe,  {\em Sov. Phys. JETP} {\bf 56}, 493(1982).
\bibitem{l} C. Itzykson and J. Zuber,  {\em Quantum Field Theory}
 (McGraw-Hill,  New York,  1980).
\bibitem{m} J. Govaerts,  L. J. Reinders,  F. De Viron,  and J. Weyers, 
{\em Nucl. Phys.} {\bf B283},  706(1987).
\bibitem{n} C. B. Chiu,  J. Pasupathy,  and S. J. Wilson, 
{\em Phys. Rev.} {\bf D32}, 1786(1985).
\bibitem{o} E. M. Henley,  WY. P. Hwang, and L. S. Kisslinger, 
{\bf Phys. Rev.} {\bf D46},  431(1992).
\bibitem{q} H. Forkel and M. K. Banerjee,  {\em Phys. Rev. Lett.} {\bf 71}
,  484(1993).
\end{thebibliography}
\end{document}